\documentclass[letterpaper, 10 pt, conference]{ieeeconf}  

\IEEEoverridecommandlockouts                              

\usepackage{caption}
\usepackage[noadjust]{cite}
\usepackage{graphicx}
\usepackage{amsmath,amssymb,thmtools,enumerate}
\usepackage{setspace}
\usepackage{booktabs}
\usepackage[caption=false,font=footnotesize]{subfig}
\usepackage{flushend} 

\newtheorem{problem}{Problem}

\newtheorem{remark}{Remark}

\newtheorem{assumption}{Assumption}

\newcommand{\m}{\mathop{\mathrm{m}}}
\newcommand{\dBm}{\mathop{\mathrm{dBm}}}
\newcommand{\Hz}{\mathop{\mathrm{Hz}}}
\newcommand{\squeezeup}{\vspace{-3mm}}

\title{\LARGE \bf
Gaussian Processes Online Observation Classification for RSSI-based Low-cost Indoor Positioning Systems
}

\author{Maani Ghaffari Jadidi, Mitesh Patel, and Jaime Valls Miro
\thanks{Maani Ghaffari Jadidi and Jaime Valls Miro are with Centre for Autonomous System, Faculty of Engineering and IT, University of Technology Sydney, Ultimo, NSW 2007, Australia {\tt\small \{maani.ghaffarijadidi, jaime.vallsmiro\}@uts.edu.au}}%
\thanks{Mitesh Patel is with FX Palo Alto Laboratory Inc., Palo Alto, CA - 94304, USA {\tt\small mitesh@fxpal.com}}%
}

\begin{document}

\maketitle
\thispagestyle{empty}
\pagestyle{empty}

\begin{abstract}
In this paper, we propose a real-time classification scheme to cope with noisy Radio Signal Strength Indicator (RSSI) measurements utilized in indoor positioning systems. RSSI values are often converted to distances for position estimation. However due to multipathing and shadowing effects, finding a unique sensor model using both parametric and non-parametric methods is highly challenging. We learn decision regions using the Gaussian Processes classification to accept measurements that are consistent with the operating sensor model. The proposed approach can perform online, does not rely on a particular sensor model or parameters, and is robust to sensor failures. The experimental results achieved using hardware show that available positioning algorithms can benefit from incorporating the classifier into their measurement model as a meta-sensor modeling technique.
\end{abstract}

\section{INTRODUCTION}
The spreading of personal communication systems into many public and private places, as well as the onset of new generation of smartphones, has enabled the development of a vast number of indoor positioning systems based on standard wireless communication technologies~\cite{liu2007,yanying2009}. While indoor radio propagation follows the same mechanisms as outdoor, shorter coverage range and greater variability of indoor environments, e.g.\@ the presence of tinted metal in windows, make modeling the radio signal attenuation significantly more challenging~\cite{rappaport1996wireless}. Further, compared to outdoor scenarios, the number of Line-Of-Sight (LOS) observations are lower which means the common Friis free space model cannot accurately model the radio signal attenuation. Therefore, for any indoor positioning system that relies on such models, the ability to differentiate LOS and Non-LOS (NLOS) observations is beneficial.

\begin{figure}
\centering
\includegraphics[width = .95\columnwidth, trim={0.5cm 0cm 1.5cm 0.25cm},clip]{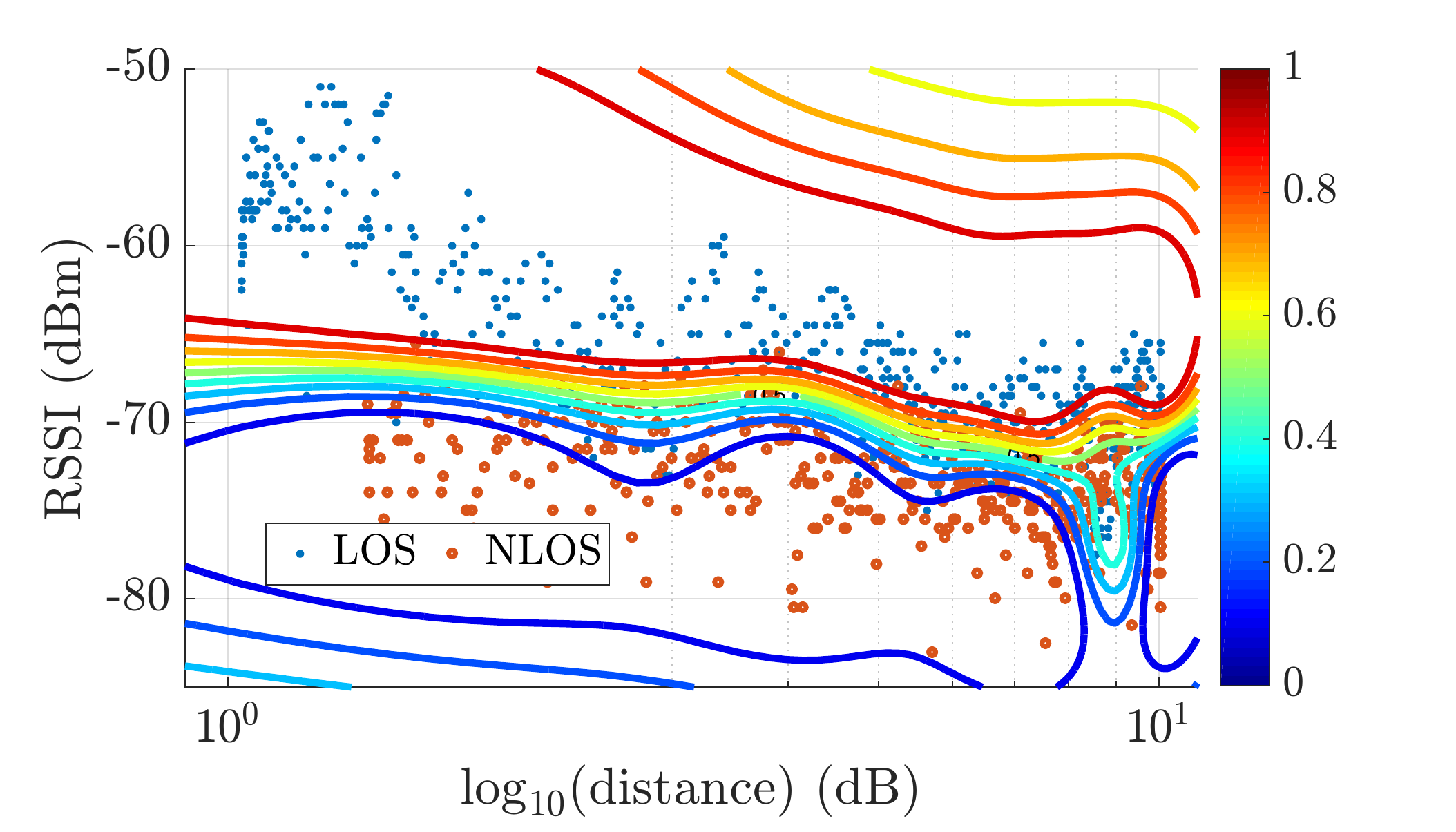}
\caption{The decision surface learned by a Gaussian process classifier using collected LOS and NLOS measurements. Each LOS/NLOS point is averaged over $6$ RSSI from $6$ co-located BLE beacons with the same transmission power. The groundtruth distances are computed using a laser range-finder sensor and an iterative closest point-based scan-matching technique.}
\label{fig:gpc}
\squeezeup
\end{figure}

In this paper, we propose a probabilistic framework to explicitly \emph{detect} and systematically \emph{mitigate} NLOS radio signal observations. The proposed approach is non-parametric, does not require a statistical characterization of waveforms, and can be incorporated into recursive Bayesian estimation frameworks such as particle filters as a meta-sensor modeling technique. We use Gaussian process classification (GPC) for offline learning of decision regions based on dense distance and Radio Signal Strength Indicator (RSSI) measurements, shown in Figure~\ref{fig:gpc}, and employ it in online scenarios using $k$d-tree structures.

\subsection{Motivation}
The main motivation stems from the challenge faced in using Bluetooth Low Energy (BLE) beacons for indoor positioning. Assuming RSSI is the only quantity available to the receiver, the common practice reported in the literature is to convert the measured RSSI to distance. However, in our experience, under realistic conditions, radio signals are severely impacted due to shadowing and multipathing effects. These incidents are due to various factors such as the presence of people, the number of reflective surfaces, and overall dynamics of the environment. Therefore, a large number of spurious measurements results in biased distance conversion and consequently poor position estimation performance. Through rejecting measurements that are not compatible with the sensor model, we only add information to the estimation process if it maintains its consistency. 

\subsection{Contributions}
The contributions of this paper are two folds. Firstly, we propose an online (adaptive) technique to model the BLE sensor so that it can tackle the shadowing and multipathing effects of the signal. Secondly, we utilize the BLE sensor model in a position estimation framework to localize a smartphone user or a robot in a given environment. It should be noted that by utilizing our approach, we eliminate the tedious process of fingerprinting the environment to generate a radio map, rather we collect data to model the BLE sensor which is a one time process and has considerably lower overhead compared to fingerprinting the environment.

\subsection{Notation}
Probabilities and probability densities are not distinguished in general. Matrices are capitalized in bold, such as in $\boldsymbol X$, and vectors are in lower case bold type, such as in $\boldsymbol x$. Vectors are column-wise and $1\colon n$ means integers from $1$ to $n$. Random variables, such as $X$, and their realizations, $x$, are sometimes denoted interchangeably. $x^{[i]}$ denotes a reference to the $i$-th element of the variable. An alphabet such as $\mathcal{X}$ denotes a set. A reference to a test set quantity is shown by $\boldsymbol x_*$. Finally, $\mathbb{E}[\cdot]$ and $\mathbb{V}[\cdot]$ denote the expected value and variance of a random variable, respectively.

\subsection{Outline}
In the following section, we present the related work. In Section~\ref{sec:problem}, the problem formulation, and required preliminaries are explained. We present details of sensor modeling and analysis in Section~\ref{sec:sensor}. The positioning algorithm is explained in Section~\ref{sec:loc}. We present the experimental results in Section~\ref{sec:result} and Section~\ref{sec:conclusion} concludes the paper.

\section{Related Work}
The idea of integrating non-parametric models into Bayesian filtering is not new. In~\cite{ko2009gp}, the system dynamics and observation models in extended and unscented Kalman filters and Particle Filters (PFs) are appropriately replaced by Gaussian Processes (GPs). In comparison to parametric models, upon the availability of sufficient training data, results show improvement in tracking accuracy. Machine learning techniques are also extensively considered for indoor localization systems. In location fingerprinting approach, kernel methods in the form of Support Vector Machines (SVMs) and GPs frameworks have become the standard way of indoor positioning~\cite{howard2003experimental,Ferris06gaussianprocesses,he2015wi,faragher2015location}. However, these approaches require the tedious process of mapping the RSSI values in different locations in the environment, prior to the experiment which is distinct from the online approach we use in this work. Furthermore, the likelihood map is non-adaptive and does not take into account dynamic of the environment.

An important part of online RSSI-based positioning systems is the radio signal path-loss model~\cite{phillips2013survey}. Such models are usually based on Friis free space model and are only valid if there is a direct and collision-free path between transmitter and receiver, with no reflection and refraction due to nearby obstacles, and in the far-field of transmitting antenna~\cite{rappaport1996wireless,goldsmith2005wireless}. A key challenge here is to be able to identify and mitigate NLOS observations~\cite{borras1998decision,gezici2003nonparametric,guvenc2007nlos,mao2007wireless,wymeersch2012machine,xiao2015non}. 
To the best of our knowledge, approaches in~\cite{wymeersch2012machine,xiao2015non} are conceptually the closest to this work. In~\cite{wymeersch2012machine}, the problem of range error mitigation using SVM and GP regression is studied. The approach uses $\ell_2$ and $\ell_1$-minimization and characterizes the ranging error based on a set of features extracted from the received waveform. In~\cite{xiao2015non}, a set of statistical features are extracted from the received signal; a classifier discards the NLOS measurements, and the distance to the transmitter is estimated using regression techniques. In this work, we do not rely on feature extraction from the received signal, the receiver has only access to the received RSSI (unlike~\cite{wymeersch2012machine}), and the classification output is incorporated into the probabilistic positioning framework for sequential estimations. In particular, instead of discarding measurements we use a probabilistic mixture measurement model. 

The technique in~\cite{meissner2014multipath} uses the floor plan to associate multipath components of the propagated radio signal to the surrounding geometry. An environment survey prior to the experiment is required as well as more sophisticated hardware for data collection. In~\cite{haneda2016indoor}, indoor channel models for a wider range of frequencies to meet 5G -- 5th generation wireless systems -- requirements are studied. The probability of LOS observations is modeled using exponential decays as a function of distance. However, it is mentioned that high variability exists between different deployments and openness of the area. It is clear that using purely distance results in a passive model and cannot cope with online radio signal variations. The proposed solution in this work is a non-parametric representation of LOS probabilities using distance and RSSI and takes the spatial correlation of radio signal propagation into account. 

\section{Problem Formulation and Preliminaries}
\label{sec:problem}
We now define the problems we study in this paper and then briefly explain the required preliminaries to solve these problems. Let \mbox{$\mathcal{M} = \{\boldsymbol m^{[j]}|j=1:n_m\}$} be a set of known and fully observable features whose elements, $\boldsymbol m^{[j]} \in \mathbb{R}^3$, represent BLE beacons locations. The robot has a receiver that can only receive the RSSI of a broadcasted signal. Let \mbox{$\mathcal{S}_t \subset \mathbb{Z}$} be the set of possible RSSI measurements at time $t$. The observation consists of an $n_s$-tuple random variable $(S_t^{[1]},...,S_t^{[n_s]})$ whose elements can take values \mbox{$\boldsymbol s_t^{[k]} \in \mathcal{S}_t$, $k \in \{1:n_s\}$}. We denote the robot position up to time $t$ by \mbox{$\boldsymbol x_{0:t} \triangleq \{\boldsymbol x_{0},...,\boldsymbol x_{t}\}$} where $\boldsymbol x_t \in \mathbb{R}^3$. Given the set of known BLE beacons and noisy observations, we wish to solve the following problems.

\begin{problem}[Measurement model]
Let \mbox{$\mathcal{Z}_t \subset \mathbb{R}_{\geq 0}$} be the set of possible range measurements at time $t$ that is calculated through a nonlinear mapping $s_t \mapsto z_t$. The measurement model $p(z_t|\boldsymbol x_t)$ is a conditional probability distribution that represents the likelihood of range measurements. Find the mapping from signal to range measurements and the likelihood function that describes the measurement noise.
\end{problem}
\begin{problem}[Positioning]
Let \mbox{$z_{1:t} \triangleq \{z_{1},...,z_{t}\}$} be a sequence of range measurements up to time $t$. Let $\boldsymbol x_t$ be a Markov process of initial distribution $p(\boldsymbol x_0)$ and transition equation $p(\boldsymbol x_t|\boldsymbol x_{t-1})$. Given $p(z_t|\boldsymbol x_t)$, estimate recursively in time the posterior distribution $p(\boldsymbol x_{0:t}|z_{1:t})$. 
\end{problem}

In the first problem, we try to characterize the received signal and through an appropriate model transform it to a range measurement. Furthermore, we need to find a likelihood function that describes the measurement noise. The second problem can be seen as a range-only self-localization problem. For simplicity, since the map is known, it is eliminated from conditional probabilities terms. We now express the main assumptions we use to solve the defined problems.

\begin{assumption}[Constant transmission power]
The transmission power of all beacons during positioning experiments remain fixed.
\end{assumption}

Since a different transmission power leads to a different signal propagation behavior, i.e.\@ a shorter or a longer range, this assumption guarantees that the sensor model complies with the employed beacons. 

\begin{assumption}[Known data association]
Each beacon has a unique hardware identifier that is available to the receiver device.
\end{assumption}

This assumption is usually satisfied in practice as each beacon has a unique MAC-address that broadcasts it together with the RSSI. Finally, we assume that the only available information to the receiver is the RSSI, this is the common case for existing wireless routers and BLE beacons. However, if the time difference of arrival (transmission time) be available to the receiver device, the position estimation accuracy can be improved.

\subsection{Bluetooth low energy technology}
Bluetooth Low Energy~\cite{bleprotocol2010} protocol was devised in 2010. It operates in the 2.4 GHz license-free band and hence shares the same indoor propagation characteristics as 2.4 GHz WiFi transceivers. Unlike WiFi, BLE uses 40 channels each with a width of 2 MHz~\cite{faragher2015}. 

\subsection{Gaussian processes classification}
Supervised classification is the problem of learning input-output mappings from a training dataset for discrete outputs (class labels). Gaussian process classification~\cite{rasmussen2006gaussian} is a non-parametric Bayesian technique that uses statistical inference to learn dependencies between points in a dataset. The problem in this paper is a binary classification. We define a training set $\mathcal{D} \triangleq \{(\boldsymbol x^{[i]},y^{[i]})|i=1:n_o\}$ of dimension $d$ which consists of a $d$-dimensional input vector $\boldsymbol x$ and a class label $y \in \{-1,+1\}$ for $n_o$ observations. In GPC, the inference is performed in two steps; first computing the predictive distribution of the latent variable corresponding to a query case, $f_*|\mathcal{D},\boldsymbol x_* \sim \mathcal{N}(\mathbb{E}[f_*],\mathbb{V}[f_*])$, and then a probabilistic prediction, $p(y_*=+1|\mathcal{D},\boldsymbol x_*)$, using a sigmoid function.

The non-Gaussian likelihood and the choice of the sigmoid function can make the inference analytically intractable. Hence, approximate techniques such as Expectation Propagation (EP)~\cite{minka2001family} needs to be used. The vector of hyperparameters (parameters of the covariance and mean functions), $\boldsymbol\theta$,  can be optimized by maximizing the log of the marginal likelihood function, $\log \ p(\boldsymbol y|\boldsymbol X,\boldsymbol\theta)$, where $\boldsymbol X$ is the $d\times n$ design matrix of aggregated input vectors $\boldsymbol x$, and \mbox{$\boldsymbol y = [y^{[1]},...,y^{[n]}]^T$}.

The GPC model implemented in this work uses a constant mean function, squared exponential covariance function with automatic relevance determination as described in~\cite{neal1996bayesian}, whereas the error function likelihood (probit regression), and EP technique for approximate inference is done using the open source Gaussian process (GP) library in~\cite{rasmussen2006gaussian}.

\subsection{Particle filters}
In the problem of localization using RSSI, the observation space is nonlinear, and the posterior density is often multi-modal. Particle filters are a non-parametric implementation of the Bayes filter that are suitable for tracking and localization problems where dealing with global uncertainty is crucial~\cite{doucet2001sequential,ristic2004beyond,thrun2005probabilistic}. In this work, we use Sample Importance Resampling (SIR) filter embedded with the systematic resampling algorithm. To detect the degeneracy and perform resampling, we compute the effective sample size which corresponds to the reciprocal of the sum of squares of particle weights.

\begin{figure*}
\centering
\subfloat[]{
    \includegraphics[width = 0.77\columnwidth, trim={0.75cm 0cm 2cm 1.cm},clip]{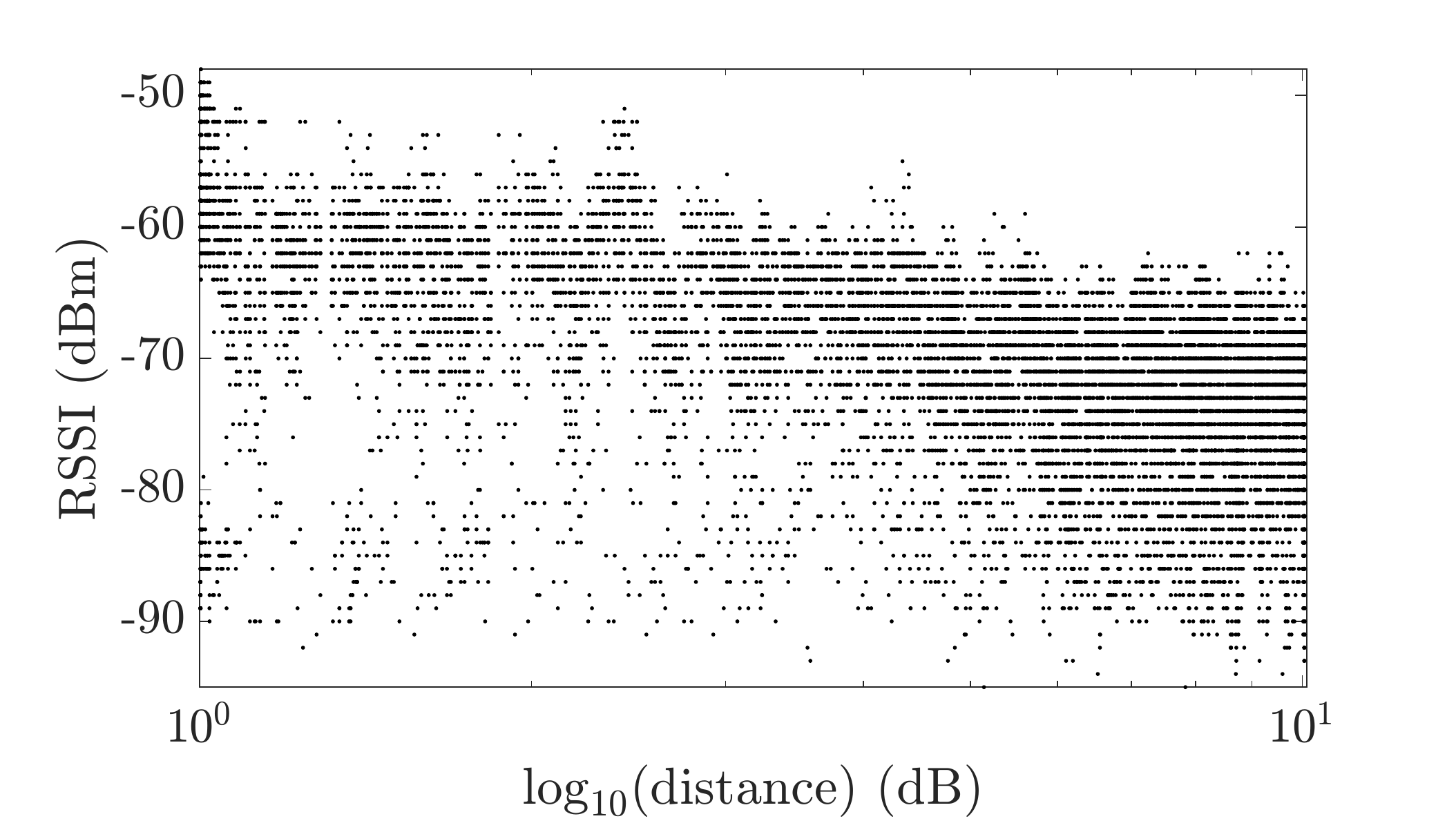}
    \label{fig:raw_los}
    }\quad
\subfloat[]{
    \includegraphics[width = 0.77\columnwidth, trim={0.75cm 0cm 2cm 1.cm},clip]{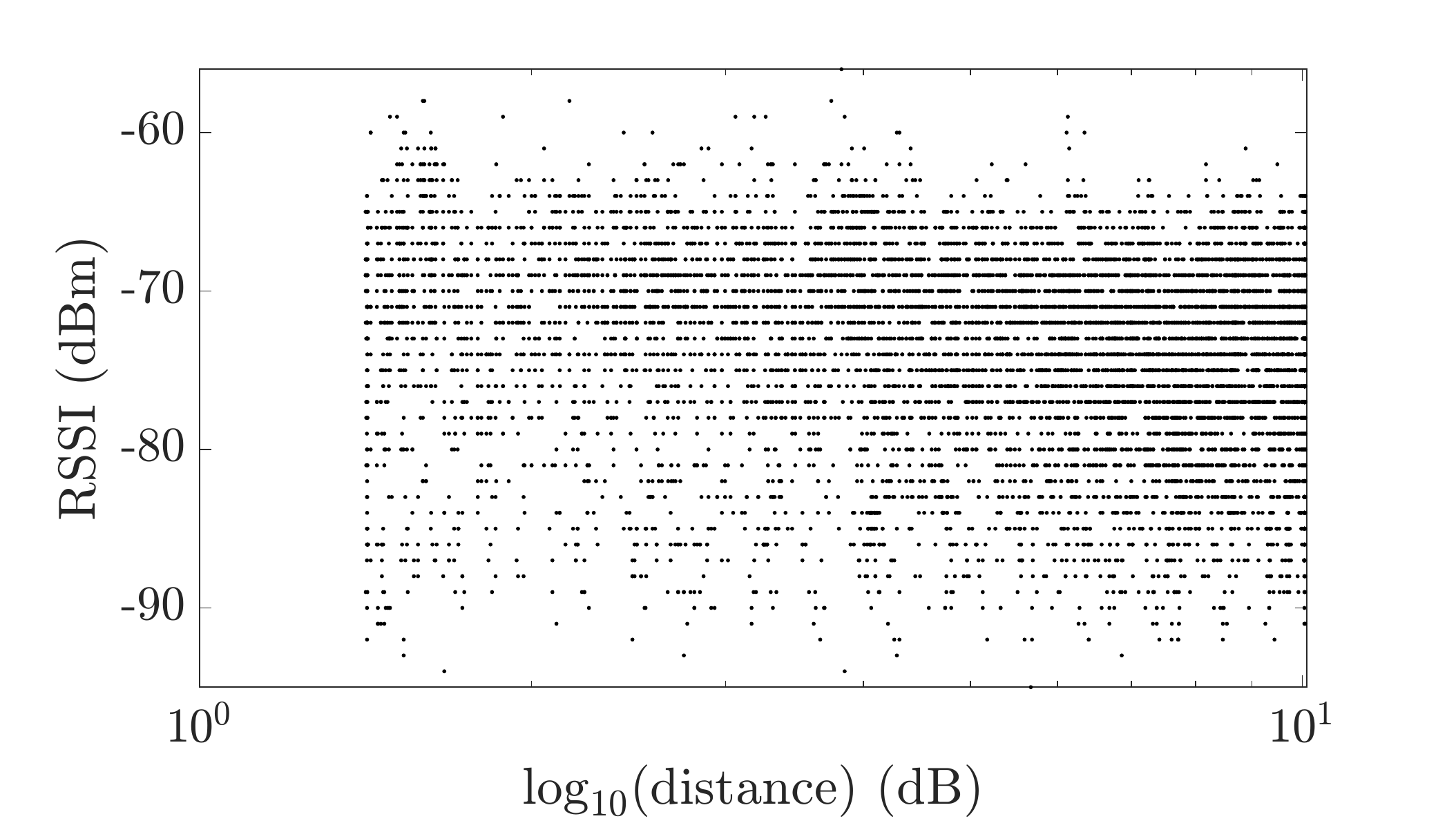}
    \label{fig:raw_nlos}
    }\\
\subfloat[]{
    \includegraphics[width = 0.77\columnwidth, trim={0.75cm 0cm 2cm 1.cm},clip]{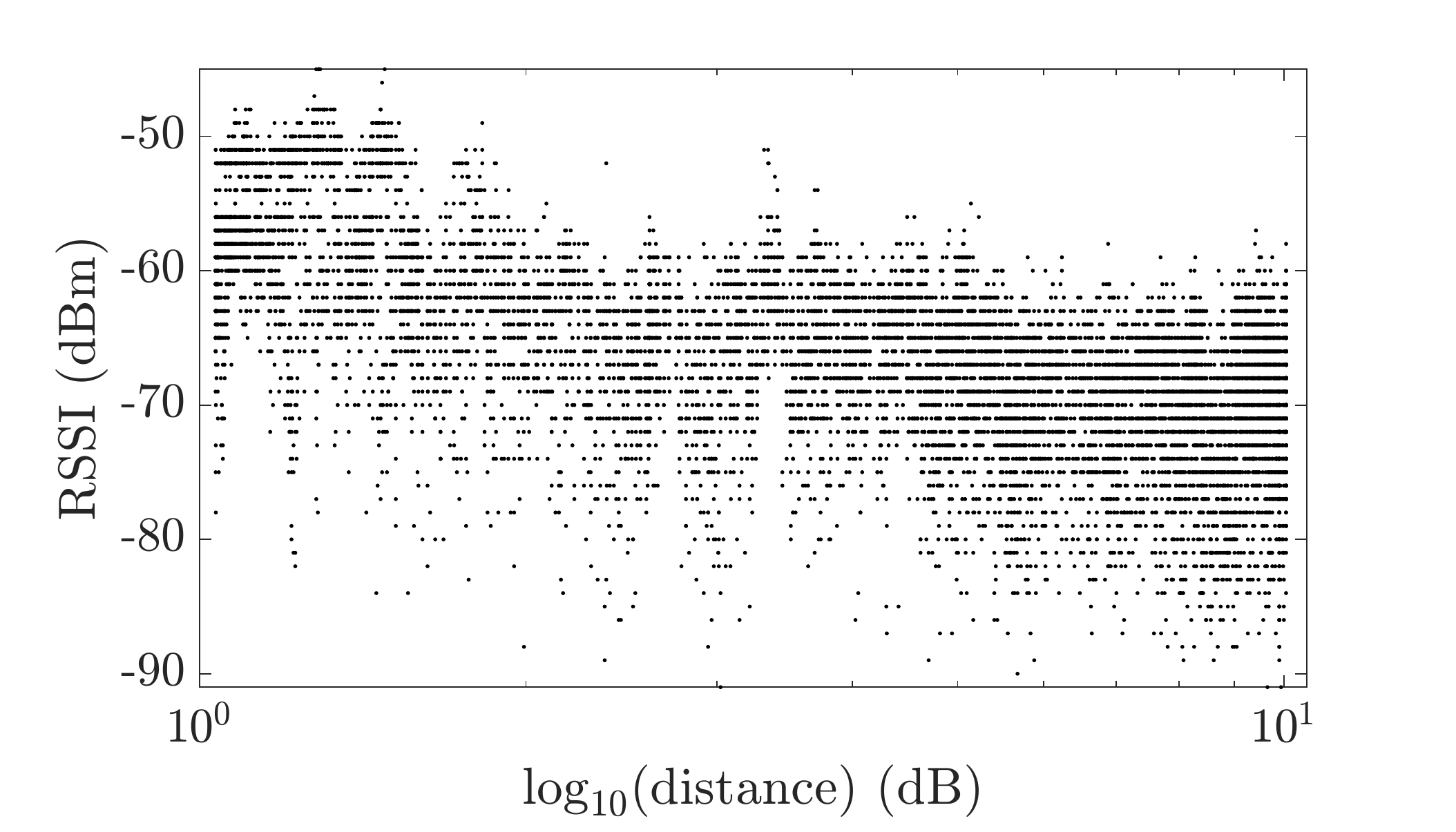}
    \label{fig:raw_los_r2}
    }\quad
\subfloat[]{
    \includegraphics[width = 0.77\columnwidth, trim={.75cm 0cm 2cm 1cm},clip]{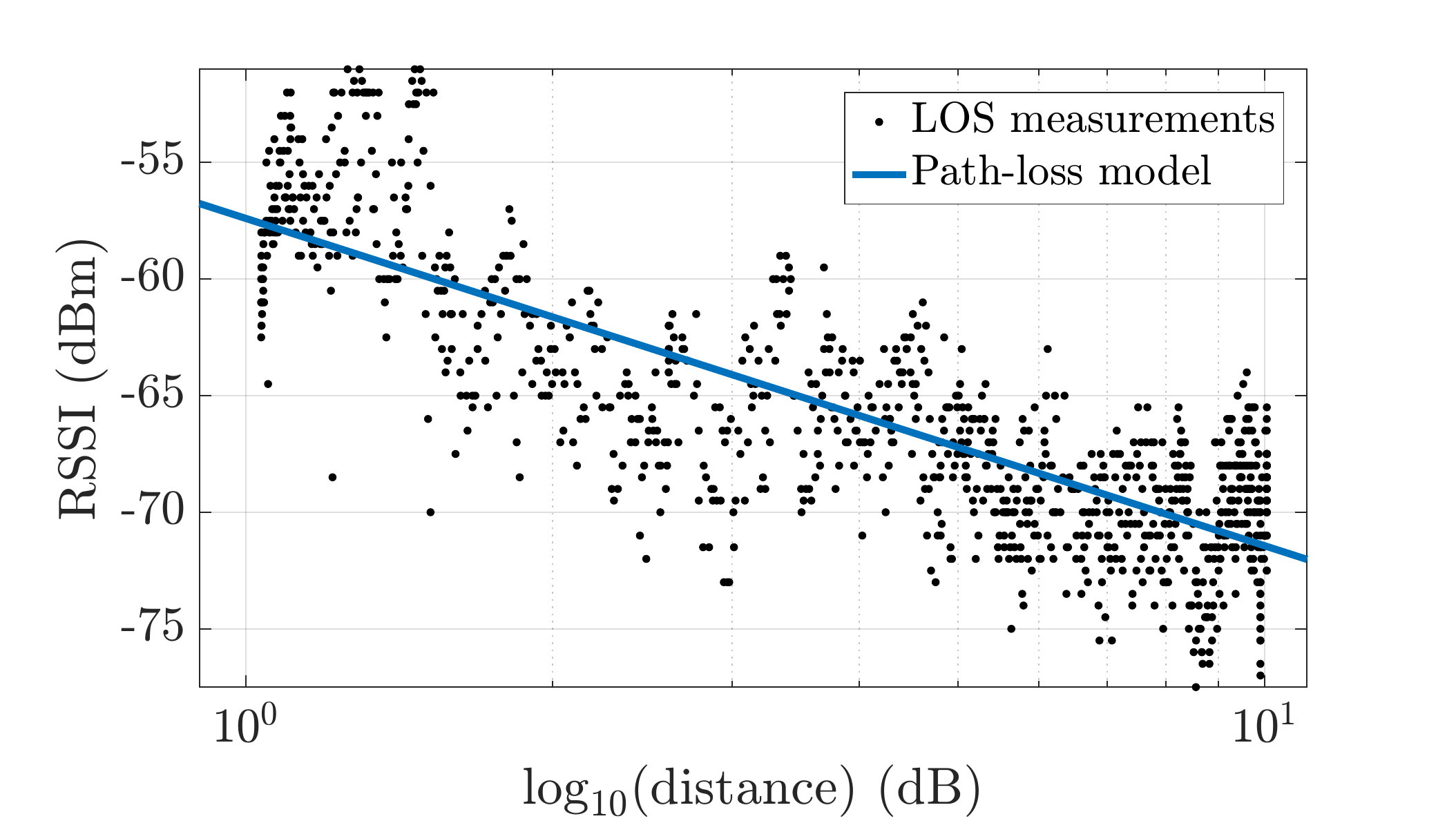}
    \label{fig:pathlossmle}
    }
\caption{Raw RSSI measurements are from $6$ co-located BLE beacons collected along $10\m$ range for (a) Round I: LOS ($12680$ points) and (b) Round I: NLOS ($9380$ points). The NLOS measurements have lower signal strength due to shadowing and non-constructive multipathing effects. (c) shows Round II: LOS ($10640$ points), and (d) shows path-loss model parameter estimation using the maximum likelihood and a Gaussian noise model. The points indicate the median of measurements from all $6$ beacons with a similar time-stamp, i.e\@ within ten milliseconds.}
\label{fig:raw_6rssi}
\squeezeup
\end{figure*}

\section{Sensor Modeling and Analysis}
\label{sec:sensor}
In this section, we tackle the first problem. To model the mapping from the signal to the measurement space, i.e.\@ RSSI to range, we use Friis free space model~\cite{rappaport1996wireless,goldsmith2005wireless} in which the signal attenuation is proportional to the logarithm of the distance. This model can characterize radio signals propagation in LOS scenarios; however, in NLOS and the presence of clutter, it may perform poorly which negatively affects the positioning algorithm. We first describe experimental data collection rounds, followed by how we use the experimental data to estimate the parameters of the path-loss model and train the GP classifier.

\subsection{Data collection rounds}
We employ a robot equipped with an Inertial Measurement Unit (IMU) and a laser range-finer to localize using laser odometry. We use this result as a proxy for groundtruth to estimate distances to the BLE beacons at known locations. 

We empirically found that the effective range of BLE beacons to define a meaningful relation between RSSI and distance is about $10\m$, which is consistent with the available literature~\cite{liu2007}. Hence, all data collection rounds for sensor modeling are performed along a $10\m$ range to capture the main trend of data. In Round I, RSSIs are collected in LOS and NLOS scenarios. The NLOS is created artificially by blocking the LOS using furniture such as chairs. In Round II, on a different working day, we collected another LOS dataset. The collected data from Round I and II are illustrated in Figure~\ref{fig:raw_6rssi}. 

\subsection{Path-loss model parameters estimation}
The signal propagation in an indoor environment is a complex physical phenomenon, and it is often not possible to find a unique model to characterize it. However, the simplified free space path-loss model can capture the essence of signal propagation. The model depends on, $a_X$ in $\dBm$, which captures the transmission power, antenna characteristics and the average channel attenuation, the received power, $p_{RSSI}$ in $\dBm$, the path-loss exponent $\gamma$, and a reference distance, $d_0$ in $\m$, for the antenna far-field. The model can be expressed as follows.
\begin{equation}
\label{eq:pathloss}
\small
p_{RSSI} = a_X + 10\gamma\log_{10}(\frac{z}{d_0}) + \epsilon
\end{equation}
where $\epsilon$ is the received signal power noise and assumed to have an independent and identically distributed (i.i.d.) Gaussian distribution, $\epsilon \sim \mathcal{N}(0,\sigma^2)$. The three model parameters $a_X$, $\gamma$, and $d_0$ can then be estimated using the nonlinear least squares parameter estimation technique, i.e.\@ maximum likelihood estimation with a Gaussian noise assumption. Figure~\ref{fig:pathlossmle} shows the model with parameters estimated using the Round II dataset.
\begin{remark}
From Equation~\eqref{eq:pathloss}, it is clear that if $p_{RSSI}$, in $\dBm$, follows a normal distribution, then the received power, in Watt, follows a log-normal distribution. Therefore, we can assume that the distance follows a log-normal distribution as well. In practice, we calculate the range $z$ from a known value of $p_{RSSI}$.
\end{remark}

\subsection{GP classifier training and validation}
\label{subsec:gpclassifier}
To increase the diversity of training data, we use NLOS observations from Round I and LOS observations from Round II. The total number of raw observations taken from $6$ BLE beacons is about $20,000$. We compute the median of the observations within ten milliseconds to reduce the effect of outliers and improve the accuracy of the training set, leading to about $2000$ points. We then downsample data to about $1000$ points to keep the computational aspect of GPC manageable. Each training point input consists of a \mbox{$2$-dimensional} vector concatenated from the RSSI observation and the corresponding groundtruth range. The target labels are set to $+1$ and $-1$ for LOS and NLOS, respectively. Figure~\ref{fig:gpc} shows the inferred probability surface in which the higher probabilities correspond to LOS observations. Note that in practice one does not have access to the groundtruth distance. Instead, the estimated distance to a beacon together with the RSSI observation are the input. To employ the classifier online, the results are stored in a $k$d-tree data structure with an appropriate resolution.

\begin{figure}
\centering
\includegraphics[width = 0.68\columnwidth, trim={0.5cm 0.75cm 1.5cm 1cm},clip]{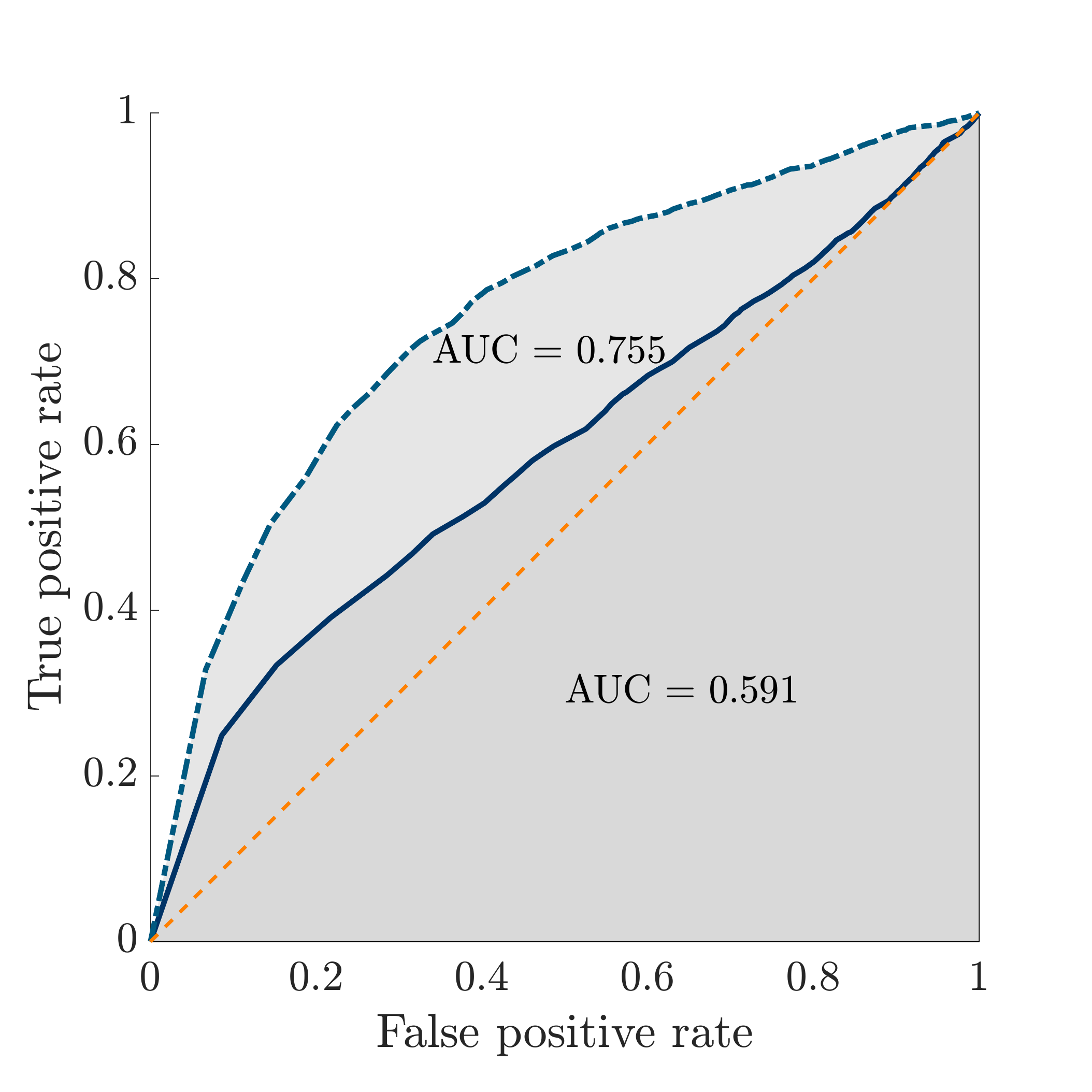}
\caption{The receiver operating characteristic curve and the area under the curve for the trained GP classifier. The classifier is validated using the LOS and NLOS measurements collected on Round I and II. The average performance of the classifier on the larger test set is lower, $0.591$.}
\label{fig:auc_gpc}
\squeezeup
\end{figure}

We evaluate the performance of the classifier using the Receiver Operating Characteristic curve (ROC) and the area under the ROC (AUC)~\cite{fawcett2006introduction}. The raw measurements without any filtering are used to conduct two tests. First, we use all observations from Round I NLOS and Round II LOS. In the second test, we use all observations from Round I and II which contain about $32,000$ points. Figure~\ref{fig:auc_gpc} illustrates the ROC analysis results where the AUC indicates the average performance of the classifier on each test set.

\section{Positioning Algorithm}
\label{sec:loc}
We now formulate a measurement model that embeds the classifier into the Bayesian filtering algorithm. Let $C^{[i]}_t$ be a Bernoulli random variable whose realization at time $t$ indicates LOS probability for the $i$-th particle. Without loss of generality, the joint probabilistic measurement model can be defined as follows.
\begin{equation}
\small
p(z_t,c^{[i]}_t|\boldsymbol x^{[i]}_t) = p(c^{[i]}_t|z_t,\boldsymbol x^{[i]}_t) p(z_t|\boldsymbol x^{[i]}_t)
\end{equation}
The conditional probability $p(z_t|\boldsymbol x^{[i]}_t)$ is the so-called likelihood function of the Bayesian filtering and in the traditional SIR filter returns an importance weight $w_t^{[i]}$ for the $i$-th particle. Therefore, the joint probability of the range measurement and LOS can be seen as a new likelihood function. However, this model is only valid if the measurement is LOS. The classifier can theoretically detect the NLOS when \mbox{$p(c^{[i]}_t|z_t,\boldsymbol x^{[i]}_t) \leq p_{los}$}, where $p_{los}$ is a threshold for LOS detection and can be set using the ROC analysis performed earlier~\cite{fawcett2006introduction}. As such, in the absence of any prior knowledge about the environment, we treat NLOS measurements as random with a constant probability $p_{rand}$. Consequently, the measurement function can be written as:
\begin{equation}
\small
\begin{split}
p(z_t,c^{[i]}_t|&\boldsymbol x^{[i]}_t) = \\ &\left\{\begin{array}{ll}p(c^{[i]}_t|z_t,\boldsymbol x^{[i]}_t) p(z_t|\boldsymbol x^{[i]}_t) & \quad \text{if}\ p(c^{[i]}_t|z_t,\boldsymbol x^{[i]}_t) > p_{los} \\
	                    p_{rand} & \quad \text{otherwise}\end{array} \right. \;
\end{split}
\end{equation}
To query the probability $p(c^{[i]}_t|z_t,\boldsymbol x^{[i]}_t)$ from the classifier, we use the raw RSSI observation and estimated distance to the corresponding beacon as
\begin{equation}
\small
h(\boldsymbol x^{[i]}_t) \triangleq \sqrt{(\boldsymbol x^{[i]}_t - \boldsymbol m^{[j]})^T(\boldsymbol x^{[i]}_t - \boldsymbol m^{[j]})}
\end{equation}
The formulated probabilistic measurement model incorporates the developed classifier into the SIR filter framework. As we will see later, one can only use $p(z_t|\boldsymbol x_t)$ to computing the filtering distribution of the robot position, e.g.\@ using a normal or a log-normal distribution, however, the joint measurement model improves the confidence about the correctness of the model-measurement relation.

As it is assumed there is no interoceptive sensor available, we do not have any knowledge regarding the transition probability model $p(\boldsymbol x_{t+1}|\boldsymbol x_{t})$. Let the state vector be \mbox{$\bar{\boldsymbol x}^{[i]}_t = [\boldsymbol x_t^{[i,1]} \dot{\boldsymbol x}_t^{[i,1]} \boldsymbol x_t^{[i,2]} \dot{\boldsymbol x}_t^{[i,2]}]^T$}, where $\dot{\boldsymbol x}^{[i]}_t$ denotes the the $i$-th particle's velocity at time $t$. Assuming a constant velocity motion model, the state equation becomes:
\begin{equation}
\small
\bar{\boldsymbol x}^{[i]}_{t+1} = \boldsymbol F \bar{\boldsymbol x}^{[i]}_t + \boldsymbol u, \quad 
\small{
\boldsymbol F = 
\begin{bmatrix}
	1 & t_s & 0 & 0 \\
	0 & 1 & 0 & 0 \\
	0 & 0 & 1 & t_s \\
	0 & 0 & 0 & 1
\end{bmatrix}}
\end{equation}
where $t_s$ is the sampling time, $\small{\boldsymbol u \sim \mathcal{N}(\boldsymbol 0, \boldsymbol Q)}$, and $\small{\boldsymbol Q}$ is a diagonal motion noise covariance matrix. Note that the receiver height installed on the robot is fixed as the robot operates on an even floor.

\begin{table}
\footnotesize
\centering
\caption{Parameters used in the positioning experiments.}
\begin{tabular}{lll}
\toprule
Parameter				& Symbol	& Value \\ \midrule
\multicolumn{3}{l}{$-$ Compared SIR particle filter variants:} \\
Gaussian   				& PFG	& -	\\
Gaussian with classifier 		& PFG-C				& -	\\  
Lognormal 				& PFL				& -	\\ 
Lognormal with classifier		& PFL-C 		& -	\\
\multicolumn{3}{l}{$-$ Path-loss model parameters:} \\
Attenuated transmission power  		& $a_X$			& -64.53			\\
The path-loss exponent 			& $\gamma$		& 1.72			\\  
Reference distance 			& $d_0$			& 1.78 $\m$	\\ 
\multicolumn{3}{l}{$-$ Measurement model:} \\
Classifier threshold			& $p_{los}$		& 0.4 	\\
Gaussian; standard deviation		& $\sigma_{n}$ 	& 3 $\m$		\\
Gaussian; random probability		& $p_{rand}$	& 0.1 	\\
Lognormal; standard deviation		& $\sigma_{ln}$ & 0.4 $\dBm$	\\
Lognormal; random probability		& $p_{rand}$	& $(d_0 \sigma_{ln} \sqrt{2\pi})^{-1}$	\\
\multicolumn{3}{l}{$-$ Motion model:} \\
Position standard deviation		& $\sigma_{u}$ 	& 0.1 $\m$		\\
Velocity standard deviation		& $\sigma_{v}$ 	& 0.05 $\m/\sec$	\\
\multicolumn{3}{l}{$-$ Particle filter:} \\
Number of particles			& $n_p$			& 100	\\ 
Resampling threshold			& $n_{thr}$		& 20	\\
\multicolumn{3}{l}{$-$ BLE Beacon Parameter:} \\
Transmission Power			& $T_x$			& +4 dbm 	\\ 
Broadcasting Frequency			& $B_{f}$		& 10 Hz	\\ \bottomrule
\end{tabular}
\label{tab:param}
\squeezeup
\end{table}

\section{Experimental Results}
\label{sec:result}
To validate the proposed measurement modeling using the GP classifier, we evaluate our approach on an indoor positioning algorithm using BLE beacons. The dataset is collected during working hours in an office space and the robot is moved with a moderate speed of $0.2~\m/\sec$ on average. In the following, we explain the experimental setup and results as well as a discussion on the limitations of this work and computational complexity analysis of the proposed algorithm. 


\subsection{Experimental setup and evaluation criteria}
Traditionally, Cram{\'e}r-Rao Lower Bound (CRLB) has been developed and used for system designs and evaluations, since it can predict the achievable performance before building the system~\cite{tichavsky1998posterior,ristic2004beyond}. We utilized CRLB to approximate the theoretical lower bound for the mean-squared error. We define the efficiency, $\eta$, of a system using $\sqrt{\text{CRLB}}$ and the Root Mean Squared Error (RMSE) as follows.
\begin{equation}
\small 
\eta = \frac{\sqrt{\text{CRLB}}}{\text{RMSE}}\times 100
\end{equation}

The explanations of the compared techniques and used parameters are provided in Table~\ref{tab:param}. We compare the results for indoor positioning using the SIR Particle Filter (PF) with Gaussian (PFG) and log-normal (PFL) likelihood functions, and with and without incorporating the classifier, \mbox{PFG-C} and \mbox{PFL-C}, respectively. To detect the degeneracy, we calculate the effective sample size, $n_{eff}=(\sum_{i=1}^{n_p}w_t^{[i]})^{-1}$, and perform resampling when $n_{eff} < n_{thr}$; where $n_p$ is the number of particles and $n_{thr}$ is a threshold $1 < n_{thr} < n_p$. All the results presented in this paper use $n_p=100$ and $n_{thr}=20$, and the robot position is estimated using the weighted average of all particles' positions. Moreover, the transmission power $T_x$ of all beacons is $+4\dBm$.

\begin{figure}[th!]
\centering
\includegraphics[width = .99\columnwidth, trim={1.8cm 3.75cm 1.6cm 4.5cm},clip]{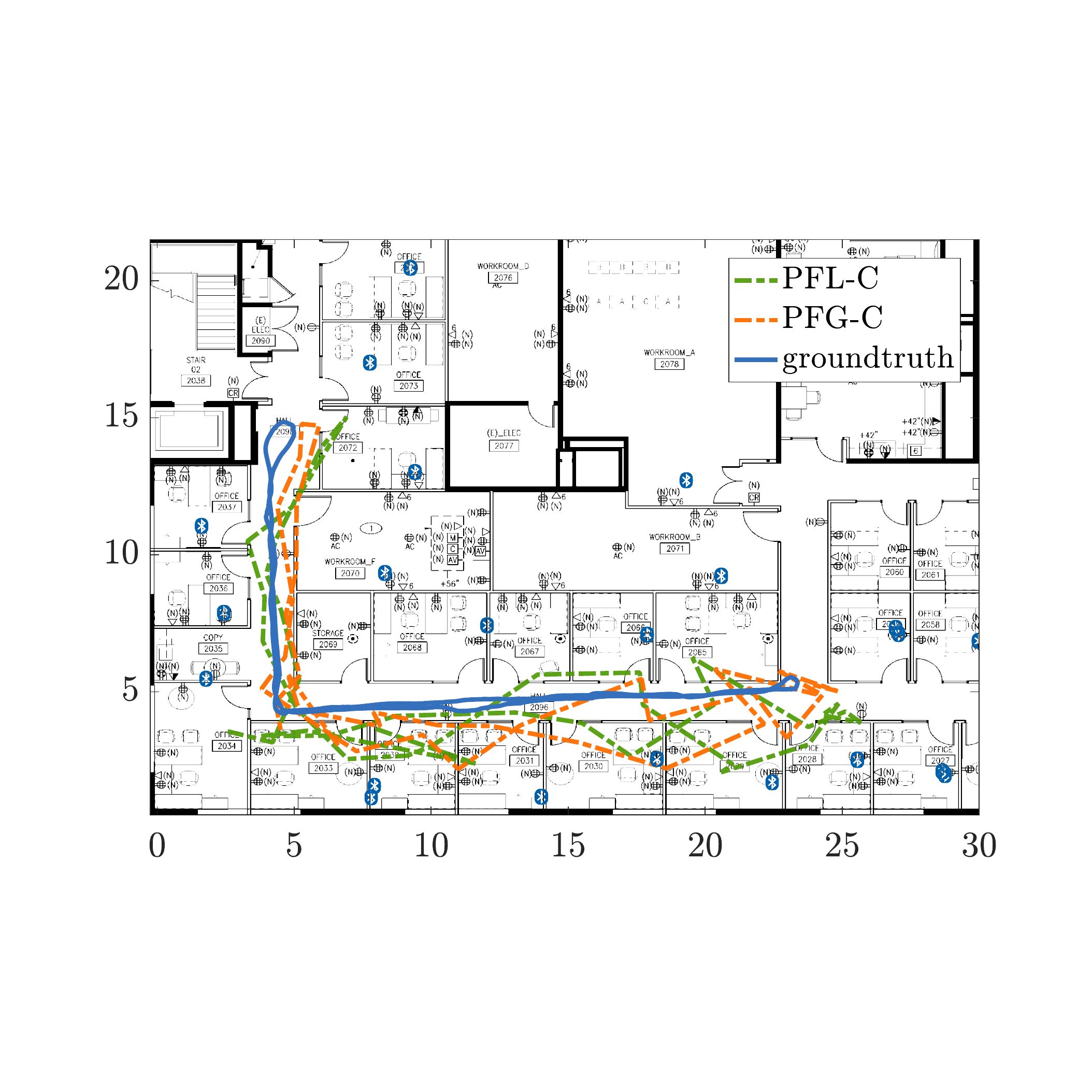}
\caption{The indoor positioning results in an office environment populated with BLE beacons. For clarity, The estimated trajectories are plotted by skipping $50$ time steps between any two successive positions.}
\label{fig:fx_office}
\end{figure}

\subsection{Indoor positioning results}
The dataset is collected in a research office partitioned into separate office cabins and consists of traditional office furniture. The data is collected using a TurtleBot equipped with an IMU sensor and a laser range-finder which are used for groundtruth pose estimation. The beacons signals are recorded using a smartphone Android app. The dataset is collected by maneuvering the robot over a distance of $70 \m$ in an office space of $20\times40 \m^2$, as shown in Figure~\ref{fig:fx_office}.

The methods are implemented using Robot Operating System (ROS)~\cite{quigley2009ros} and results for indoor positioning are processed using MATLAB. The nominal sampling rate is BLE beacons is $10\Hz$; however, in practice, we experienced a sampling rate of $7\Hz$, on average, for the entire dataset.

\begin{table}[t!]
\footnotesize
\centering
\caption{Comparison of indoor positioning algorithms using particle filtering with and without incorporating the online classifier on Dataset I and II. The results are averaged over $100$ runs; mean $\pm$ standard error.}
\resizebox{\columnwidth}{!}{
\begin{tabular}{lcccc}
\toprule
				& PFG &	PFG-C & PFL & PFL-C \\ \midrule
 
$\sqrt{\text{CRLB}}$ ($\m$)	& $0.4254$		& $0.4254$ 		& $0.0747$		& $0.0747$ 	\\ 
RMSE ($\m$)			& $8.08\pm 0.38$	& $1.99\pm 0.01$ 	& $4.11\pm 0.06$	& $3.06\pm 0.05$ 	\\
$\eta$ (\%)			& $5.76\pm 0.12$	& $21.36\pm 0.13$ 	& $1.85\pm 0.02$	& $2.50\pm 0.04$ 	\\
Time ($\sec$)			& $10.6\pm 0.01$	& $182.7\pm 0.05$ 	& $12.0\pm 0.02$ 	& $180.8\pm 0.29$ 	\\ \bottomrule
\end{tabular}}
\label{tab:casdataset}
\squeezeup
\end{table}

\begin{figure}[t!]
\centering
\includegraphics[width = 0.8\columnwidth, trim={0.5cm 0cm 1.5cm 0cm},clip]{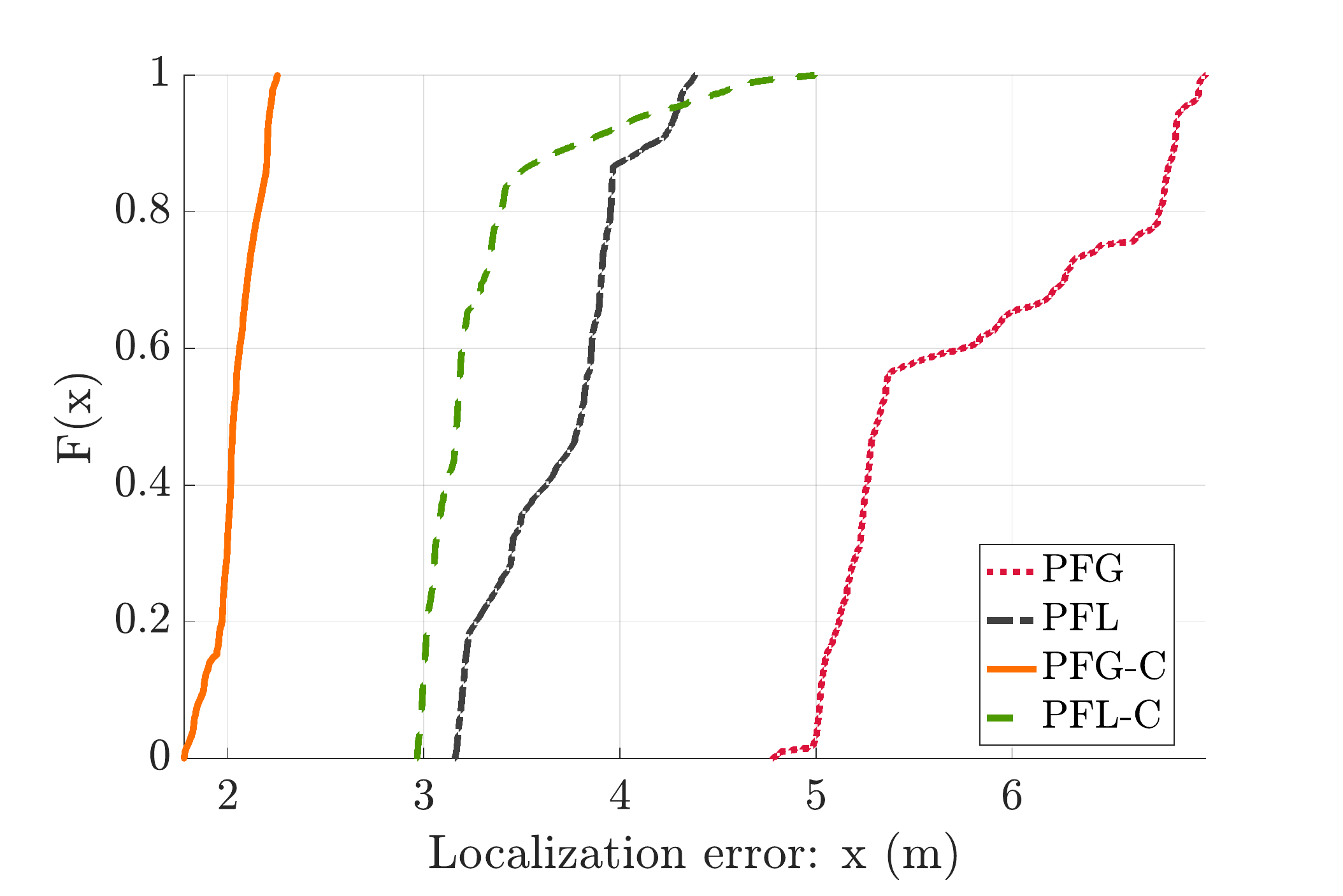}
\caption{The empirical cumulative distribution functions of the four compared techniques.}
\label{fig:fx_cdf}
\squeezeup
\end{figure}

Figure~\ref{fig:fx_cdf} shows the empirical cumulative distribution function (CDF) of the four compared techniques. The empirical CDF is an unbiased estimate of the population CDF and is a consistent estimator of the true CDF. Each curve illustrates the median of $100$ CDF from $100$ independent runs. The PFG-C demonstrates the best performance by the localization error of about $2\m$. Note that faster rise from zero to one along the vertical axis is a desirable outcome.

\begin{figure}[t]
\centering
\includegraphics[width = 0.6\columnwidth, trim={0.25cm 1.5cm 1.5cm 1cm},clip]{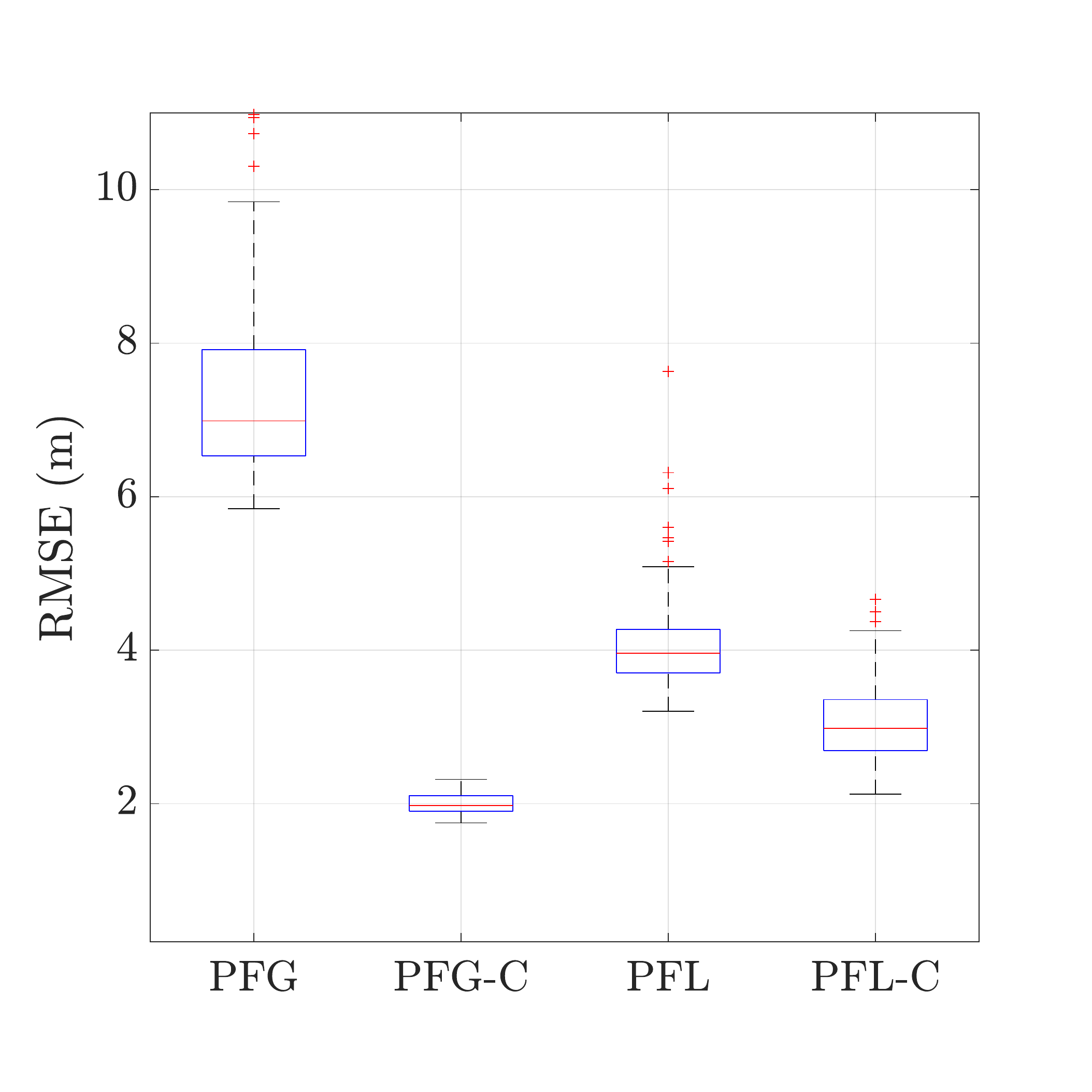}
\caption{The statistics from indoor positioning results using particle filtering with normal and log-normal noise distributions. The incorporation of the classifier into the sensor model leads to a more accurate location and scale estimation.}
\label{fig:fx_box}
\squeezeup
\end{figure}

The statistical summary of the results is depicted in Figure~\ref{fig:fx_box}. As an example, the estimated trajectory using PFG-C and PFL-C are also illustrated in Figure~\ref{fig:fx_office}. The proposed classifier has a desirable effect on the robot position estimation where the robot position has fewer fluctuations. The classifier makes the positioning algorithm more robust to noisy observations and outliers, improving the overall reliability of the system (Figure~\ref{fig:fx_cdf}). This is, in particular, appealing for the case of the normal likelihood. From the physical nature of the radio signal propagation, the ranging bias is always positive. Therefore, a symmetric distribution such as the Gaussian likelihood performs poorly in characterizing the noise. However, depending on the parameters, there are instances that the normal and log-normal distributions behave similarly. Nevertheless, the classifier improves the estimation performance for both types of noise models.

Table~\ref{tab:casdataset} shows the numerical comparison between different algorithms from $100$ independent runs. The CRLB value for normal and log-normal distributions is inherently different as the noise variance for the former is in meters and the latter in $\dBm$. Thus, one should compare the efficiency of methods with a similar likelihood function. However, we can compare all algorithms using RMSE. Overall, PFG-C and PFL-C show better performance compared to their corresponding algorithms that do not use the classifier.

\subsection{Discussion}
The main limitation of the proposed online classification technique is that the RSSI range varies according to the BLE beacon TP. Therefore, using beacons that have different TPs as compared to those used for training the classifier will result in lower performance. Furthermore, the classifier cannot improve the sensor model if it is not (at least empirically) compatible with the underlying physical nature of the RF signal propagation. Therefore, it can only act as a proxy for consistent observation selection which can detect and mitigate destructive multipathing, shadowing, or sensor failures, i.e.\@ weak batteries or hardware failures.

Finally, to our experience, collection of NLOS data is of great importance. If NLOS data has a substantial overlap with LOS data, then the performance of the trained classifier will decrease dramatically. In small environments, this effect can be understood from constructive multipathing or partially blockage of LOS during data collection.

\subsection{Computational complexity}
For $n_o$ observations, the approximate inference using EP scales as $O(n_o^3)$ which is performed offline. The $k$d-tree structure is suitable for efficient search in low-dimensional spaces, such as the case in this work. For $n_p$ particles and $n_z$ nearest neighbor queries, the algorithm scales as $O(n_p n_z \log n_q)$, where $n_q$ is the number of stored query points, and usually $n_z \ll n_p$.

\section{Conclusion and Future Work}
\label{sec:conclusion}
We studied the problem of indoor positioning using BLE beacons. We developed an online classification strategy to improve the consistency of received measurements with the employed sensor model. Our experimental results under realistic conditions show promising improvements and the proposed classifier can be used as a meta-sensor modeling technique to cope with spurious measurements. The proposed method is particularly simpler and more scalable than the popular fingerprinting technique as the training phase is in the sensor space instead of spatial coordinates of an environment.

The future work includes further studies and improvement of the sensor model in the presence of semi-dynamic obstacles. Integration of incremental motion measurements such as IMUs can also improve the accuracy of position tracking. Moreover, increasing the sampling rate can provide a better efficiency through a higher flow of information into the estimation process. Lastly, the simultaneous estimation of the robot (receiver) and BLE beacons positions is an interesting avenue to follow.


\section*{ACKNOWLEDGEMENT}
\label{sec:acknowledgement}
\small{
This work has been supported by Yahoo Research under the Faculty Research and Engagement Program (FREP) an academic outreach initiative. The authors would also like to thank FX Palo Alto Laboratories Inc., for sharing Bluetooth Low Energy dataset collected by them to validate our proposed algorithm.}

\addtolength{\textheight}{-12cm}   

\bibliographystyle{IEEEtran} 
\bibliography{refs}

\end{document}